\documentclass[preprint,aps,superscriptaddress]{revtex4}

\usepackage{graphicx}
\usepackage{dcolumn}
\usepackage{bm}

\begin{document}

\preprint{APS/123-QED}

\title{Nodeless energy gaps of single-crystalline Ba$_{0.68}$K$_{0.32}$Fe$_{2}$As$_{2}$ as
    seen via  $^{75}$As NMR}

\author{Z. Li}
\affiliation{Institute of Physics and Beijing National Laboratory for Condensed Matter Physics, Chinese Academy of Sciences, Beijing 100190, China}

\author{D. L. Sun}
\affiliation{Max Planck Institute - Heisenbergstrasse 1, D-70569 Stuttgart, Germany}

\author{C. T. Lin}
\affiliation{Max Planck Institute - Heisenbergstrasse 1, D-70569 Stuttgart, Germany}

\author{Y. H. Su}
\affiliation{Institute of Science and Technology for Opto-electronic Information, Yantai University, Yantai 264005, China}

\author{J. P. Hu}
\affiliation{Department of Physics, Purdue University, Indiana 47907, USA}
\affiliation{Institute of Physics and Beijing National Laboratory
for Condensed Matter Physics, Chinese Academy of Sciences, Beijing
100190, China}


\author{Guo-qing Zheng}%
\affiliation{Department of Physics, Okayama University, Okayama 700-8530, Japan}
\affiliation{Institute of Physics and Beijing National Laboratory for Condensed Matter Physics, Chinese Academy of Sciences, Beijing 100190, China}

\date{\today}

\begin{abstract}
We report $^{75}$As nuclear magnetic resonance studies on a very clean hole-doped single-crystal Ba$_{0.68}$K$_{0.32}$Fe$_{2}$As$_{2}$ ($T_{\rm {c}}=38.5$
K). The spin-lattice relaxation rate $1/T_{1}$ shows an exponential
decrease below $T \simeq 0.45 T_{\rm c}$ down to $T \simeq 0.11
T_{\rm c}$, which indicates a fully-opened energy gap. From the ratio
$(T_{1})_{c} / (T_{1})_{a}$, where $a$ and $c$ denote the crystal
directions, we find that the antiferromagnetic spin fluctuation is
anisotropic in the spin space above $T_{\rm c}$. The anisotropy
decreases below $T_{\rm c}$ and disappears  at $T \rightarrow 0$.
We argue that the anisotropy stems from spin-orbit coupling whose
effect vanishes when  spin-singlet electron pairs form with a nodeless
gap.

\end{abstract}

\maketitle


The discovery of superconducting transition in electron-doped iron-arsenide
LaFeAsO$_{1-x}$F$_{x}$ provides a new route to high temperature
superconductivity \cite{YKamihara}. Remarkably,
many other $R$FeAsO$_{1-x}$F$_{x}$ ($R$: rare earth)
were synthesized and $T_{\rm {c}}$ was raised to 55 K in
SmFeAsO$_{1-x}$F$_{x}$ \cite{ZARenSm},  which is the highest among
materials except cuprates. Soon after these works, the hole-doped
BaFe$_{2}$As$_{2}$ was also found to be superconducting \cite{122}.
The large single crystals of Ba$_{1-x}$K$_{x}$Fe$_{2}$As$_{2}$ are
easy to obtain, which makes them  a good system for studying many
physical quantities.

One of the most outstanding issues for a new superconductor is the
symmetry of  the electron pairs which is directly
related to the paring mechanism. Nuclear magnetic resonance
(NMR) experiments found the electron pairs to be in the spin-singlet
state \cite{KMatanoPr} and indicated the existence
of multiple energy gaps  \cite{KMatanoPr, SKawasaki}. The multiple-gap
property is likely associated with the multiple electronic bands.
The Fermi surfaces  consist of two hole-pockets centered at the
$\Gamma$ point and two electron pockets around the
$M$ point \cite{DJSingh}. However, whether there are nodes in the gap function
or not is still under hot debate. Angle-resolved photoemission
spectroscopy (ARPES) \cite{HDing}
suggested fully opened gaps, but thermal conductivity \cite{LiSL,Taillefer}
measurements suggested nodal gaps.
The penetration depth measurements by different groups  have led to opposite conclusions \cite{KHashimoto,
CMartin}.

Theoretically, the sign-reversing
$s^{\pm}$-wave model has been considered as the most promising
candidate \cite{IIMazin, kuroki, fwang}, but $d$ wave or $s$ wave with zero gap, and even a conventional $s^{++}$ wave was also
proposed \cite{SGraser, Thomale,Kontani}. It has been
shown that the $s^{\pm}$-wave or a multiple-gap $d$-wave model can fit
quite well the spin-lattice relaxation rate, $1/T_{1}$, which shows
a rapid decrease below $T_{\rm c}$ with a hump structure at $T \sim
T_{\rm c}/2$ \cite{KMatanoPr,SKawasaki, YBang, parish, KMatanoBa}. However, an
important  feature that $1/T_{1}$ should decrease as an
exponential function of $T$ expected for the $s$-wave gaps
has not been observed so far, because of impurity scattering in the
samples. The impurity scattering can also alter other physical
properties \cite{Johnston}. Thus, the conclusions on
the gap symmetry drawn so far are still controversial. Measurements
in sufficiently clean samples are highly needed to resolve the issue.

Here we report $^{75}$As NMR study on a very clean single crystal Ba$_{0.68}$K$_{0.32}$Fe$_{2}$As$_{2}$ with $T_{c}=38.5$ K that is the highest among reports for this
family.
We obtained two pieces of evidence for fully-opened gaps. First, we observe an exponential decay of  $1/T_{1}$ below $T\simeq 0.45 T_{\rm c}$ down to $T \simeq 0.11 T_{\rm c}$. For the second piece of evidence, we find that the antiferromagnetic (AF) spin fluctuation (SF) is anisotropic in the spin space above $T_{\rm c}$, but the anisotropy decreases below $T_{\rm c}$ and disappears at $T \rightarrow 0$. We argue that the anisotropy is due to spin-orbit coupling, whose effect vanishes at $T \rightarrow 0$ because the electron pairs are in the spin-singlet state with nodeless gap.

The single crystal of Ba$_{0.68}$K$_{0.32}$Fe$_{2}$As$_{2}$ was
grown by using the self-flux method and characterized as discussed elsewhere \cite{GLSun}.
Both dc susceptibility measured by a superconducting quantum interference device and ac
susceptibility measured by the NMR coil indicates $T_{\rm{c}} =38.5
$ K at zero magnetic field. The  $T_{\rm{c}}$ is 37.6 K for $\mu_{0}
H$ (=7.5T) $\parallel a$ axis and 36.4 K for $\mu_{0} H$ (=7.5T) $
\parallel c$ axis.
 The  $1/T_{1}$ was
determined from an excellent
fitting to
$1-M(t)/M(\infty)=0.1$exp$(-t/T_{1})+0.9$ exp$(-6 t/T_{1})$, where $M(t)$ is the nuclear
magnetization at time $t$ after the saturation pulse \cite{ANarath}.

\begin{figure}
\includegraphics[width=6.5cm,clip]{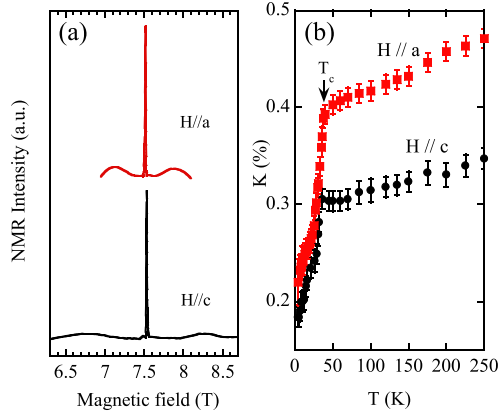}
\caption{\label{fig:spectra} (Color online) (a) $^{75}$As-NMR spectra at a frequency of $\omega_0/2 \pi = 55.1$ MHz and $T$=100 K. The vertical axis for $H \parallel a$ is offset for clarity. (b) The $T$ dependence of the Knight shift with $H \parallel a$ axis and $H \parallel c$ axis, respectively. The arrow indicates $T_{\rm c}$ for $H \parallel a$.}
\end{figure}

Figure \ref{fig:spectra} (a) shows the $^{75}$As-NMR spectra by scanning the magnetic field at a fixed frequency, $\omega_0/2 \pi = 55.1$ MHz. The nuclear quadrupole frequency $\nu _Q$ is found to be  5.1 MHz at 100 K which is  smaller than that in the Sn-flux-grown sample (5.9 MHz) \cite{KMatanoBa}. Since doping of K increases  $\nu _Q$ \cite{KMatanoBa}, this suggests that the Sn-flux-grown crystal had a higher doping rate. The Knight shift  $K$ was obtained from the central transition peak and
determined with respect to $\omega_0 / \gamma$ with  the nuclear gyromagnetic ratio $\gamma =7.2919$
MHz/T. Below $T_{\rm c}$,  $K$ is  obtained by scanning $\omega_0$ at a fixed field to avoid the vortex pinning effect. Above $T_{\rm c}$, we confirmed that the results obtained by scanning field and scanning frequency agree well. The effect of the nuclear quadrupole interaction was taken into account in extracting $K_{a}$.
As shown in Fig. \ref{fig:spectra} (b), both $K_{a}$ and $K_{c}$ show a sharp decrease below $T_{\rm c}$, which  indicates spin-singlet pairing \cite{KMatanoPr}.

\begin{figure}
\includegraphics[width=5.5cm,clip]{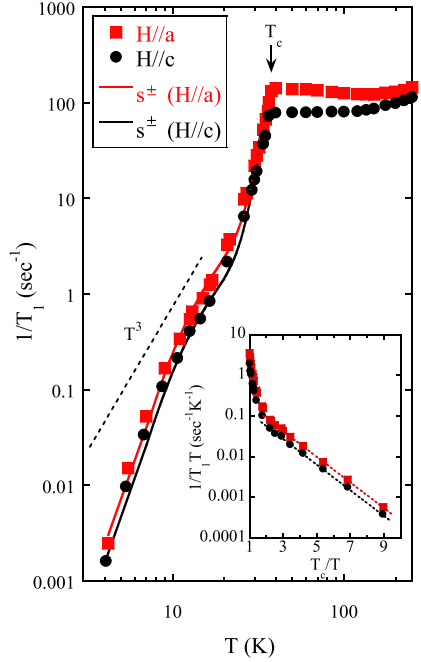}
\caption{\label{fig:T1} (Color online) The $T$ dependence of $1/T_{1}$. The error is within the size of the symbols. The dashed line shows the $T^3$ variation.
The curves below $T_{\rm{c}}$  are fits to a two-gap $s^{\pm}$ model using the same parameters for both directions. The inset shows the semilog plot of $1/T_{1}T$ vs $T_{\rm c}/T$, which evidences an activation type $T$ dependence of the relaxation..}
\end{figure}

The main panel of Fig. 2
shows the $T$ dependence of $1/T_{1}$ that decreases rapidly below $T_{\rm c}$, with the reduction over about five decades. The decrease  at low $T$ is much faster than $T^3$ that is expected for a $d$-wave gap. As in other materials, $1/T_{1}$ shows a ``knee'' shape around half $T_{\rm{c}}$, which indicates multiple gaps \cite{KMatanoPr, SKawasaki}. To see the low-$T$ behavior more clearly,  we plot $1/T_{1}T$ as a function of inverse reduced-temperature $T_{\rm{c}}/T$ in the inset.
As can be seen there, $1/T_{1}T$ shows a very good exponential behavior below 17 K. This is  strong evidence for a fully opened gap.


Using the $s^{\pm}$-wave model and introducing the impurity scattering rate $\eta$ in the energy spectrum, $E=\omega + i \eta$ \cite{ZLi}, but neglecting the quasiparticle damping effect for simplicity, we can fit the data quite well. For a sign-reversing two-gap model, as seen in Fig. 2, we obtain $\Delta_{1}^{+}= 5.63$ $k_{\rm{B}} T_{\rm{c}}$, $\Delta_{2}^{-}=1.11$ $k_{\rm{B}} T_{\rm{c}}$, $N_{1} : N_{2} =0.85 : 0.15$, where $N_{i}$  is the density of state (DOS) on band $i$, and $\eta = 0.044$ $k_{\rm B}T_{\rm c}$.
For  a model of three bands corresponding to ARPES \cite{HDing}, we obtain  $\Delta_{1}^{+}= 4.7$ $k_{\rm{B}} T_{\rm{c}}$, $\Delta_{2}^{+}=0.96$ $k_{\rm{B}} T_{\rm{c}}$, $\Delta_{3}^{-}= 4.7$ $k_{\rm{B}} T_{\rm{c}}$, $N_{1} : N_{2} : N_{3}=0.44 : 0.12 : 0.44$, and $\eta = 0.022$ $k_{\rm B}T_{\rm c}$.
The  $\eta$ is much smaller than $\eta=0.15$ $k_{\rm B}T_{\rm c}$ in LaFeAsO$_{0.92}$F$_{0.08}$ \cite{SKawasaki} and $\eta=0.22$ $k_{\rm B} T_{\rm c}$ in the Sn-flux grown Ba$_{0.72}$K$_{0.28}$Fe$_{2}$As$_{2}$  \cite{KMatanoBa}, meaning that the present sample is much cleaner, as supported by a small resistivity of 26$\mu$$\Omega$ at $T_c$ \cite{GLSun} and the much sharper  spectrum-width that is only half  the value for the Sn-flux grown crystal. The cleanness of the present crystal is the  reason for the  exponential behavior of $1/T_{1}$ at low $T$; impurity scattering brings about finite DOS that results in seemingly power-law $T$-dependence of $1/T_{1}$   \cite{KMatanoPr, ZLi, Grafe,Fukazawa,Yashima}. It should be emphasized that the coherence peak is not seen even in such clean sample, which seems hard to be explained by a $s^{++}$-wave gap.

\begin{figure}
\includegraphics[width=5.5cm,clip]{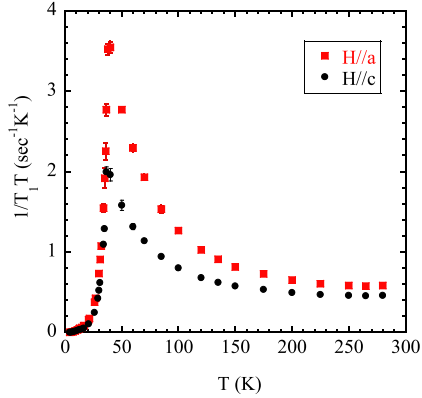}
\caption{\label{fig:T1T} (Color online) $T$-dependence  of the $^{75}(1/T_{1}T)$. The arrow indicates $T_c$.}
\end{figure}

\begin{figure}
\includegraphics[width=5.5cm,clip]{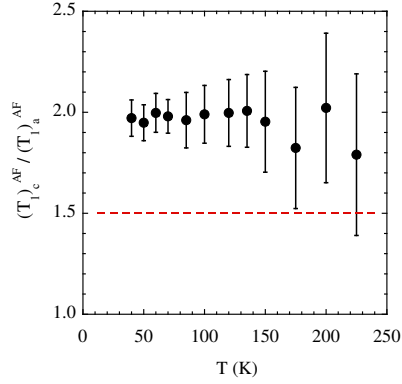}
\caption{\label{fig:RAF} (Color online) $T$-dependence of the  anisotropy of $T_{1}$ due to AF spin fluctuation. The dashed  line marks the value for  isotropic AF spin fluctuation.}
\end{figure}

Next we move to the normal state. Figure \ref{fig:T1T} shows the $T$ dependence of $1/T_{1}T$ which increases with decreasing $T$ down to $T_{\rm c}$, indicating strong AF SF. The $1/T_{1}T$ stems from
the magnetic susceptibility at  all wave vectors. When there exists strong AF SF, one may assume $1/T_{1}T=(1/T_{1}T)^{\rm AF}+(1/T_{1}T)^{0}$, where $(1/T_{1}T)^{\rm AF}$ is due to the  susceptibility at the AF wave vector $Q$, and $(1/T_{1}T)^{0}$ is due to $s$-band electrons and the orbital hyperfine interaction. Note that above $T=250$K, $1/T_{1}T$ becomes a constant. By taking the averaged value of $1/T_{1}T$ at $T \geq 250$K as $(1/T_{1}T)^{0}$, $(1/T_{1}T)^{\rm AF}$ is then obtained. Figure \ref{fig:RAF} shows the ratio of the relaxation  due to AF SF, $(T_{1})_{c}^{\rm AF}/(T_{1})_{a}^{\rm AF}$, which is about $2$. This result indicates that the SF is anisotropic in the spin space, as elaborated below.

Generally, $1/T_{1}$ is related to the transverse fluctuating internal magnetic field, $\delta H$, as follows \cite{TMoriya}:
\begin{eqnarray}
\left(\frac {1}{T_{1}}\right)_{z}=&\frac{\gamma^2}{2}\int _{-\infty}^{\infty} {\rm d}t {\rm cos}(\omega_{0}t)
&\left<  \delta H_{x}(t) \delta H_{x}(0) + \delta H_{y}(t) \delta H_{y}(0) \right>
\label{eq:T1T},
\end{eqnarray}
where $\left<\cdots\right>$ denotes the statistical average. $\delta {\bf H}$ is related to the fluctuating moment ${\bf S}$ of Fe as $\delta {\bf H}={\bf A}\cdot{\bf S}$, where ${\bf A}$ is the hyperfine coupling tensor between the As nucleus and Fe spins.

For $Q=(\pi, 0)$ and $Q=(0, \pi)$ AF SF, one has \cite{KKitagawa}
\begin{eqnarray}
{\bf A}(\pi,0)
=
\left(%
\begin{array}{ccc}
  0 & 0 & A \\
  0 & 0 & 0 \\
  A & 0 & 0 \\
 \end{array}%
\right)
,\ \ {\rm and}\ \
{\bf A}(0,\pi)
=
\left(%
\begin{array}{ccc}
  0 & 0 & 0 \\
  0 & 0 & A \\
  0 & A & 0 \\
 \end{array}%
\right)
\label{eq:T1a},
\end{eqnarray}
respectively.
One therefore obtains
\begin{eqnarray}
(1/T_{1})_{a,b}^{\rm AF} & = & \frac{\gamma^{2}}{4}A^{2}\int_{-\infty}^{\infty}{\rm d} t {\rm cos}(\omega_{0}t) \left< S_{a}(t)S_{a}(0)+S_{b}(t)S_{b}(0)+S_{c}(t)S_{c}(0) \right>,\\
(1/T_{1})_{c}^{\rm AF}  & = & \frac{\gamma^{2}}{2}A^{2}\int_{-\infty}^{\infty}{\rm d} t {\rm cos}(\omega_{0}t) \left<S_{c}(t)S_{c}(0) \right>
\label{eq:T1b}.
\end{eqnarray}
Since $\left<S_{j}(t)S_{j}(0)\right>$ $(j=a,b,c)$ can be expressed in terms of  the imaginary part of the  susceptibility $\chi ''_{j}$ through the fluctuation-dissipation theorem,
$\int_{-\infty}^{\infty}{\rm d} t {\rm cos}(\omega_{0}t) \left< S_{j}(t)S_{j}(0) \right> = \frac{2k_{\rm B}T\chi''_{j}(\omega_{0})}{(\gamma_{e}\hbar)^{2}\omega_{0}}$
, the anisotropy of the relaxation  can be expressed as
\begin{eqnarray}
R_{\rm AF}=\frac{(1/T_{1})_{a}^{\rm AF}}{(1/T_{1})_{c}^{\rm AF}} =  \frac{\chi''_{a}(\omega_{0},Q)+\chi''_{b}(\omega_{0},Q)}{2\chi''_{c}(\omega_{0},Q)}+\frac{1}{2}
\label{eq:RAF}.
\end{eqnarray}
If $\chi''_{a}(\omega_{0},Q)=\chi''_{b}(\omega_{0},Q)=\chi''_{c}(\omega_{0},Q)$, namely, if the SF is isotropic in the spin space, then
\begin{eqnarray}
R_{\rm AF}^{\rm iso}=1.5
\label{eq:RAF2}.
\end{eqnarray}
The observed $R_{\rm AF}$ shown in Fig. \ref{fig:RAF} is much larger than 1.5, which follows from Eq. (5) that  $\chi ''_{a,b}(\omega_{0},Q)$ is larger than $\chi ''_{c}(\omega_{0},Q)$ by about 50\%.

We propose that the anisotropy in the SF stems from spin-orbit coupling (SOC) that mixes spin and orbital freedoms so that the magnetic susceptibility bears  some orbital character, which is  anisotropic. We use a two-band model \cite{SRaghu} involving
spin-orbit coupled $d_{xz}$ and $d_{yz}$ and calculate the anisotropy. Our theoretical study starts from a
two-dimensional Hamiltonian:
\begin{eqnarray}
H = H_{TB} + H_{LS} .
\end{eqnarray}
Here $H_{TB} = \sum_{\mathbf{k}\mu \nu\sigma} \varepsilon_{\mu \nu}
d^{\dag}_{\mathbf{k}\mu\sigma} d_{\mathbf{k}\nu\sigma}$ is the
tight-binding Hamiltonian with orbit index $\mu,\nu = 1(xz), 2(yz)$
and spin index $\sigma=+1 (\uparrow), -1 (\downarrow)$.  The
SOC is described by $H_{LS}=2\lambda_{LS}\sum_{l}
\mathbf{L}_{l}\cdot \mathbf{S}_{l}
=-i\lambda_{LS}\sum_{\mathbf{k}\sigma}\sigma
d^{\dag}_{\mathbf{k},1\sigma} d_{\mathbf{k},2\sigma} + h.c.$ .


The dynamical magnetic response is calculated  at the
random-phase-approximation level, with the on-site intra-orbit
Hubbard interaction
$H_{U} =\sum_{l\in Lattice,\mu} U n_{l\mu\uparrow} n_{l\mu\downarrow}$.
The longitudinal (transverse) susceptibility $\bar{\chi}_{c}$ ($\bar{\chi}_{+-}$)
 at
$\mathbf{q}$=$\mathbf{Q}=(\pi, 0)$ or $(0,\pi)$ is
\begin{eqnarray}
\bar{\chi}_{c(+-)}(\mathbf{q},i\nu_{n})=\frac{1}{1-\bar{\chi}_{0,c(+-)}(\mathbf{q},i\nu_{n})
\Gamma(\mathbf{q}) }\bar{\chi}_{0,c(+-)}(\mathbf{q},i\nu_{n}) ,
\end{eqnarray}
where $\bar{\chi}$ is a $2\times 2$ matrix  in the orbital space
with the matrix elements defined by
$\bar{\chi}_{c}^{\mu \nu}(\mathbf{q},\tau)= \langle T_{\tau} S^{z}_{\mu}
(-\mathbf{q},\tau) S^{z}_{\nu}(\mathbf{q},0) \rangle$ and  $\bar{\chi}_{+-}^{\mu \nu}(\mathbf{q},\tau)
= \frac{1}{2}\langle T_{\tau} S^{+}_{\mu} (-\mathbf{q},\tau)
S^{-}_{\nu}(\mathbf{q},0) \rangle $.
$\Gamma(\mathbf{q})=\text{diag}(U,U)$ is the Hubbard interaction
vertex in the spin-spin channel.
The bare  susceptibility
$\bar{\chi}_{0}(\mathbf{q},i\nu_{n})$ can be easily obtained by
diagonalizing the   Hamiltonian shown in Eq. (7).
We choose the nearest-neighbor hopping integral  $t_{1}=-0.1051$ eV \cite{More2009}.

The calculated magnetic anisotropy above $T_c$ with
$\lambda$=0.2$|t_{1}|$ ($U=9.7|t_{1}|$) and
$\lambda$=0.1$|t_{1}|$ ($U=9.87|t_{1}|$), leading to  $R_{AF}\sim$2
at $T=T_c$, is respectively shown in Fig. 5, which is in qualitative agreement with the
experimental finding. Here $T_c$=0.065$|t_{1}|$ is obtained by a
self-consistent calculation of a mean-field BCS model with a gap
$\Delta=0.309|t_{1}|$.


\begin{figure}
\includegraphics[width=6cm,clip]{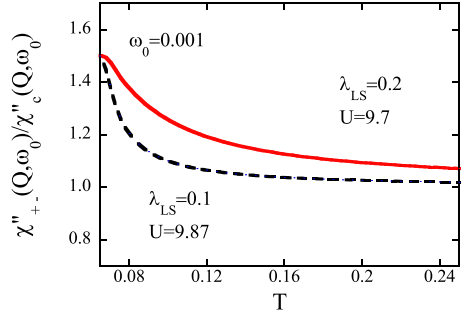}
\caption{\label{fig:Lambda} (Color online) Calculated magnetic anisotropy above $T_c$ due to SOC. The parameters and $k_BT$ are in units of $|t_{1}|$, where $t_{1}$ is the nearest-neighbor hopping integral.}
\end{figure}

\begin{figure}
\includegraphics[width=6cm,clip]{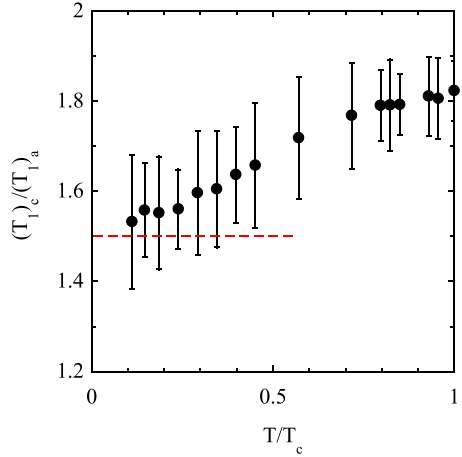}
\caption{\label{fig:R} (Color online) $T$-dependence of the $T_{1}$ anisotropy below $T_{c}$. The dashed straight line indicates the value for isotropic AF SF.}
\end{figure}

A particular feature we find experimentally is that the $T_{1}$ ratio decreases below
$T_{\rm c}$ and it approaches the  characteristic value $1.5$ for
the isotropic SF. Below $T_{\rm c}$, it is less trivial to subtract
the contribution of $(1/T_{1}T)^{0}$, so we simply plot the raw data
as shown in Fig. \ref{fig:R}. We emphasize, however, that this
approximation does not affect our conclusion \cite{footnote}, since
the contribution from $(1/T_{1}T)^{0}$ to the observed $1/T_{1}T$ is
only  $15\%$ for $H \parallel a$ and $20\%$ for $H \parallel c$ at
$T=T_{\rm c}$.

The asymptotic value  $1.5$  for $(T_{1})_{c}/(T_{1})_{a}$
implies that the SOC effect vanishes
 at $T \rightarrow 0$. This can happen only when the gaps
are fully opened in the case of spin-singlet pairing with
$\Delta > \lambda_{\rm LS}$. When there are nodes in the gap
function, $1/T_{1}$ at low
$T$ is governed by the nodal quasiparticles  that are spin-orbit coupled, thereby $(T_{1})_{c}/(T_{1})_{a}$
should resume its value of $2.0$ at $T = T_{\rm c}$.
Note also that $(T_{1})_{c}/(T_{1})_{a}$ should become $\sim$0.75 if the AF SF completely vanishes \cite{KKitagawa}.
Thus, our finding of the decrease of $(T_{1})_{c}/(T_{1})_{a}$ to $1.5$ at
$T \rightarrow 0$ is another strong evidence for nodeless gap and implies that the AF SF persists in the superconducting state.

In conclusion, from the NMR measurements on a clean single crystal
Ba$_{0.68}$K$_{0.32}$Fe$_{2}$As$_{2}$,  we find the long-sought exponential
decrease of $1/T_1$ at low $T$,
which evidences a fully-opened gap. In the normal state, the AF SF is anisotropic in the spin space. However, the
anisotropy diminishes below $T_{\rm c}$ and vanishes at the zero-$T$ limit, which is a feature indicating nodeless
gap.

We thank S. Kawasaki,  K. Matano and M.  Ichioka for help, and I. Eremin, Z. Fang, H. Ikeda and Z.-Y. Lu  for useful discussions. This work  was supported by CAS, research grants from  JSPS  and MEXT, and NSFC No. 10974167 (YHS).



\begin{thebibliography}{}
\bibitem{footnote}
Below $T_{\rm c}$, if we assume that  $(1/T_{1}T)_{0}$ follows a function as shown by the solid curve in Fig. \ref{fig:T1} and subtract it from $1/T_{1}T$, then $R_{_{ AF}}$ decreases from 2 at $T=T_{\rm c}$ to $1.58$ at the lowest temperature $T=0.11T_{\rm c}$.
\end{thebibliography}

\end{document}